\documentclass[namedreferences]{solarphysics}

\usepackage[hyperref,optionalrh,natbib]{spr-sola-addons} 
\usepackage{graphicx}        
\usepackage{color}           
\usepackage{breakurl}        
\usepackage{multirow}
\usepackage{txfonts}

\newcommand{\farcs}{\hbox{$.\!\!^{\prime\prime}$}}
\newcommand{\arcsec}{\hbox{$^{\prime\prime}$}}

\newcommand{\etal}{{\it et al.}}

\newcommand{\aap}{    {\it Astron. Astrophys.}}

\newcommand{\apj}{    {\it Astrophys. J.}}
\newcommand{\apjl}{   {\it Astrophys. J. Lett.}}

\newcommand{\mnras}{  {\it Mon. Not. Roy. Astron. Soc.}}
\newcommand{\nat}{    {\it Nature}}

\newcommand{\pasj}{   {\it Pub. Astron. Soc. Japan}}

\newcommand{\solphys}{{\it Solar Phys.}}

\begin{document}

\begin{article}

\begin{opening}

\title{Evolution of small-scale magnetic elements in the vicinity of granular-size swirl convective motions}

\author{S.~\surname{Vargas Dom\'inguez}$^{1}$\sep
        J.~\surname{Palacios}$^{2}$\sep
        L.~\surname{Balmaceda}$^{3,4}$\sep
        I.~\surname{Cabello}$^{5}$ and   V.~\surname{Domingo}$^{5}$
       }
\runningauthor{Vargas Dom\'inguez \etal}
\runningtitle{Small-scale magnetic elements in the vicinity of granular-size swirl convective motions}

   \institute{$^{1}$ Big Bear Solar Observatory, NJIT, 40386 North Shore Lane, Big Bear City, CA 92314-9672, U.S.A.
                     email: \url{svargas@bbso.njit.edu} \\ 
              $^{2}$ Space Research Group-Space Weather, Departamento de F\'isica. Universidad de Alcal\'a, 28871 Alcal\'a de Henares. Madrid. SPAIN\\
              $^{3}$ICATE/CONICET-UNSJ, CC 49, 5400 San Juan, Argentina\\
               $^{4}$INPE, P.O. Box 515, CEP 12227-010, Sao Jose dos Campos, Brazil\\
              $^{5}$ Image Processing Laboratory, Universidad de Valencia, P.O. Box 22085, E-46071, Valencia, Spain\\
               }

\begin{abstract}
Advances in solar instrumentation have led to a widespread usage of time series to study the dynamics of solar features, specially at small spatial scales and at very fast cadences. Physical processes at such scales are determinant as building blocks for many others occurring from the lower to the upper layers of the solar atmosphere and beyond, ultimately for understanding the bigger picture of solar activity. Ground-based (SST) and space-borne (Hinode) high-resolution solar data are analyzed in a quiet Sun region displaying negative polarity small-scale magnetic concentrations and a cluster of bright points observed in G-band and Ca \textsc{ii}~H images. The studied region is characterized by the presence of two small-scale convective vortex-type plasma motions, one of which appears to be affecting the dynamics of both, magnetic features and bright points in its vicinity and therefore the main target of our investigations. We followed the evolution of bright points,  intensity variations at different atmospheric heights and magnetic evolution for a set of interesting selected regions.  A description of the evolution of the photospheric plasma motions in the region nearby the convective vortex is shown, as well as some plausible cases for convective collapse detected in Stokes profiles.
\end{abstract}
\end{opening}

\section{Introduction}
     \label{S-Introduction} 

For many years, the spatial-resolution constraints in solar observations have prevented detailed observations of small-scale events in the solar atmosphere.  We are currently entering a stage of widespread usage of time series to study the dynamics of solar features at small spatial scales and at very fast cadences.  The study of multiple phenomena at such scales has proved to play a major role in understanding the physical processes that take place at different heights, from the lower to the upper layers of the solar atmosphere. Photospheric plasma motions are, in particular, of great interest as they are directly affecting the footpoints of magnetic loops rooted underneath the visible surface and generating deformations (i.e. stresses, twisting and others) that might end up changing the topology of magnetic field lines on their way up from the photosphere to the chromosphere and beyond. Recent observations suggest that a substantial part of the magnetic flux in quiet Sun regions might be the result of the emergence of these small-scale and short-lived magnetic loops \citep[see][and references therein]{mjmartinez2009,palacios2012} and that can also contribute to heating of the solar chromosphere and corona \citep{wang1995,li2007}.
 
Convective process take place as a mechanism of energy exchange involving plasma dynamics at different spatial and temporal scales. Small-scale convectively driven vortex-type motions have been discovered by \cite{bonet2008} using high-resolution observations whilst tracking bright points (hereafter BPs) in the solar photosphere. Theory predicted these types of motions but only recently, and thank to the highly resolved solar images, that these whirlpools have lately been detected. Moreover, these authors established logarithmic spiral trajectories followed by the BPs while being swallowed by a downdraft (i.e. sink). Vortex flows are commonly found in simulations at the vertices between multiple granules \citep{nordlund1986,danilovic2010}. Evidences of photospheric vortex flows have been also observed at larger scales \citep{brandt1988} such as supergranular junctions \citep[see also the paper by][and references therein]{attie2009}. A recent study by \cite{wedemeyer2009} has revealed swirl events, featuring dark and bright fast rotating patches in time series of chromospheric quiet Sun regions inside coronal holes close to disk center. Though the motion of BPs seems to be connected to the presence of these swirl events, the actual physical link is yet to be found. 

\citet{danilovic2010} observed strong downflows as consequence of the intensification of magnetic field in the context of a convective collapse, and these downflows helped the creation of small vortices. Some authors theorize about the idea that the turning motion of BPs leads to magnetic braiding and the field lines can act as guide lines to spread waves from the photosphere to the chromosphere \citep{Jess2009,Goode2010}.  ~\citet{kitiashvili2011} simulated the turning motion and the subsequent appearance of waves.  More recently,\citet{wedemeyer2012} proposed that such vortex motions can act as energy channels into higher layers in the solar atmosphere through the comparison of multi-wavelength observations and MHD simulations.  The sizes of these events also vary, depending on the spatial scale of the study. The size of the vortex was more than 7~Mm and it persisted for several hours in \citet{attie2009}. These eddies were found in a variety of solar scales, from supergranular sizes  to the medium-sized ones \citep{brandt1988, balmaceda2010}, to the smallest ones that are detected and tracked using G-band bright points \citep[GBPs,][]{bonet2008, bonet2010} with a size of 0.5~Mm. The smallest vortices from simulations are presented in \citet{danilovic2010}, with a diameter of 0.2 Mm. The values for the vorticity are usually around $\sim$ 10$^{-3}$ when related to observations \citep{vargas2010, bonet2010}; however, simulations usually yield higher values of one order of magnitude \citep[i.e., ][]{danilovic2010, pandey2012}. \citet{moll2011} present a simulation studying lifetimes, spatial coverage and inclinations, and their dependence to divergence. Not only quiet region GBPs have been studied but also in sunspots, as rotating umbral dots \citep{Bello2012}, or even vertical ones that may create lanes on granules \citep{Steiner2010}.

In the present work we analyze a quiet Sun region in which \cite{balmaceda2010}, hereafter Paper I, have reported on a small-scale vortical motion that appears as affecting the dynamic of magnetic concentrations. These authors described the vortex motion in the solar atmosphere associated with the rotation and \emph{engulfment} of magnetic features as evidenced by the analysis of solar magnetograms. Hereafter we will understand the words \emph{vortex} and \emph{swirl} as referred to this particular event unless otherwise stated. We have detected the presence of another convective vortex motion of similar size. Centers of divergence of both swirls are separated by a distance of only 5\arcsec. Our main interest is a multi-wavelength characterization of the region of interest (hereafter ROI) affected by vortex found by \cite{balmaceda2010}. Intense activity of BPs is identified in the region and these bright features are found to be affected by the swirling motion, as well as their associated magnetic concentrations. We study the appearance of intergranular lanes and their layouts are interpreted to be ruled by the formation of the vortex. Furthermore, we aim at describing the evolution of the photospheric plasma motions in the region as well as changes in the underlying magnetic field. We identify the formation of the vortex in the computed flow maps, and register the variation of the mean velocity values for various stages in the evolution of the convective swirl event.  We believe that these aspects (i.e. activity of BPs, magnetic concentrations, configuration of intergranular lanes  and photospheric plasma flows) are all intrinsically related though the direct connections are yet to be found by adding more observational evidence and throughout numerical simulations that can accomplish these new scenarios.  

The paper is structured as follows. Images acquisition and data processing are detailed in Section~\ref{S:2}. In Section~\ref{S:3} we present a general description of the BPs, magnetic concentrations and  intergranular lanes in the region displaying the convective vortex motion. In Section~\ref{S:4} we focused on the photospheric plasma flows to evidence the formation and evolution of the vortex. In Section~\ref{S:5} we describe the behavior of small magnetic elements under the influence of the vortex motion. Section~\ref{S:6} is finally devoted to comments and a general discussion.

\begin{table*}
\centering
\caption{Characteristics of the time series acquired from ground-based and satellite facilities.}
\begin{tabular}{clcccccc}\\
Telescope & Obs.  & Series & Time & Duration & N. images & Cadence  & FOV\\
 &   & \# &  UT &   {\footnotesize[min]} &  & {\footnotesize[sec]} &  {\footnotesize[\arcsec]}\\\hline
\multirow{2}{*}{SST} & \multirow{2}{*}{G-band} & 1 & 08:47-09:07 & 19 & 77 & 15 & 68.5~$\times$~68.5  \\
& & 2 & 09:14-09:46 & 32 & 129 & 15 &  68.5~$\times$~68.5 \\\hline\\
\multirow{4}{*}{Hinode} & CN & 1 & 08:40-09:20 &  40 & 70& 35 & 19.2~$\times$~74.1 \\
& Ca \textsc{ii}~H & 1 & 08:40-09:20 &  40 & 71 & 35 & 19.2~$\times$~74.1 \\
& MgI & 1 & 08:40-09:20 & 40 & 143 & 20 &15.4~$\times$~65.3 \\
& SP & 1 & 08:20-09:44 &  84 & 2558 scans & 2 & 2.7~$\times$~40.6  \\\hline
\end{tabular}
\label{table1}
\end{table*}
 
\section{Observations and data preparation} 
      \label{S:2}

The campaign performed at the Swedish 1-m Solar Telescope \citep[SST,][]{SST} was carried out during September-October 2007 as part of an international collaboration involving different institutions and researchers from Europe and Japan. This was a long campaign (24 days) with coordinated observations using not only the SST but also three other solar telescopes  (DOT, VTT and THEMIS) at the Canary Islands Observatories in La Palma and Tenerife, respectively. The observations represented the first joint campaign making use of all the above mentioned ground-based solar facilities and the space solar telescope Hinode \citep{kosugi2007} in the framework of the \emph{Hinode Operation Program 14}.

\subsection{Ground-based SST data} 
  \label{S:obsSST}

 The data from the SST analyzed in the present work were acquired during a particular observing run on 29 September 2007. The main target of interest was a  quiet region close to the solar disc center ($\mu$=0.99). Images in G-band ($\lambda 430.56$ nm) were recorded at a fast cadence so that the \emph{Multi-Frame Blind-Deconvolution} \citep[MFBD,][]{lofdahl1996,lofdahl2002MFBD} restoration technique could be applied to correct the data for atmospheric turbulence and instrumental aberrations that degrade the quality of the images. CCD cameras with a size of $2048 \times 2048$ square pixels were employed and the effective field-of-view (hereafter FOV) corresponded to $68 \times 68$ square arcsec  with a sampling of $0.034$\arcsec/pix. Once the observing run was over, only the best quality images were preserved for further operations. Processing of the images included first the standard flat fielding and dark current subtraction, as well as the removal of hot and dark pixels and spurious borders in the flat-fielded images. After performing these steps, the images of the sequence were grouped in sets of about 125 frames acquired within time intervals of 15 seconds each. Every set yielded one restored image almost reaching the diffraction limit of the telescope ($\sim0.1\arcsec$). Time series of restored images were finally compensated for diurnal field rotation, rigidly aligned, corrected for distortion and subsonic filtered to eliminate p-modes and residual jittering \citep{title1986}. The final product were two time series (\emph{s1} and \emph{s2}) of images (with a time gap of about 6 min between them) as listed in Table~\ref{table1} (SST).

\subsection{Hinode data}
\label{30sep}

The data from Hinode on 29 September 2007 corresponds to the coordinated observations previously described. The Solar Optical Telescope \citep[SOT,][]{tsuneta2008} onboard the Hinode satellite acquired filtegrams in the CN headband ($\lambda 388.35$ nm) and the core of Ca \textsc{ii}~H ($\lambda 396.85$ nm), with a cadence of $\sim$35~s and pixel size of 0.054\arcsec, using the Broadband Filter Imager (BFI). The observed FOV corresponds to 19.18\arcsec $\times$ 74.09\arcsec and almost the whole area of the Hinode FOV was covered by  SST observations. Table~\ref{table1} (Hinode) summarizes the parameters of the time series in more detail.  The Narrowband Filter Imager (NFI) was employed to obtain magnetograms in the Mg I line ($\lambda 517.3$ nm) at a cadence of $\sim$20~s. A sequence of images from the Spectropolarimeter (SP) instrument \citep{ichimoto2008,tsuneta2008} were also acquired. This data set comprises the full Stokes parameters: I, Q, U, V measured along a slit of 256 pixels in raster scan mode of 18 scans during 08:20--09:44 UT, with a sampling of $0.15$\arcsec/pix and FOV of $2.66\arcsec~\times~40.57\arcsec$. A  \textit{dynamic mode} was set to allow the study of extremely dynamic events by operating with an exposure time of 1.6 s per slit position. Noise level was $1.6 \times 10^{-3} I_c$ for Stokes I and $1.8 \times 10^{-3} I_c$ for Stokes Q and U. Magnetic field maps are obtained using the peak Stokes V and dopplershifts are inferred for the Stokes I. The SOT images were corrected for dark current, flat field and cosmic rays by using the standard IDL \emph{SolarSoft} routines. A subsonic filtering procedure was applied to get rid of high frequency oscillations.

\section{Description of the quiet Sun in the vortex's vicinity}
\label{S:3}

Quasi-simultaneous data taken by Hinode and SST were used. Images were co-aligned and trimmed in order to get the same FOV in all wavelengths.  In the present work we will focus in a region with a considerable concentration of GBPs that are co-spatial with negative-polarity magnetic fragments. The magnetic patches are mostly distributed along a curved path with a larger negative core in one end. This is the region studied in Paper I, where these authors found a rotation suffered by the magnetic concentration, as commented in the previous section. The rotation is accompanied by small-scale processes of fragmentation and coalescence of BPs clearly discernible in G-band filtergrams, that take place along the intergranular lanes as will be described in Section~\ref{S:5}.

\begin{figure}
\includegraphics[angle=0,width=1.\linewidth]{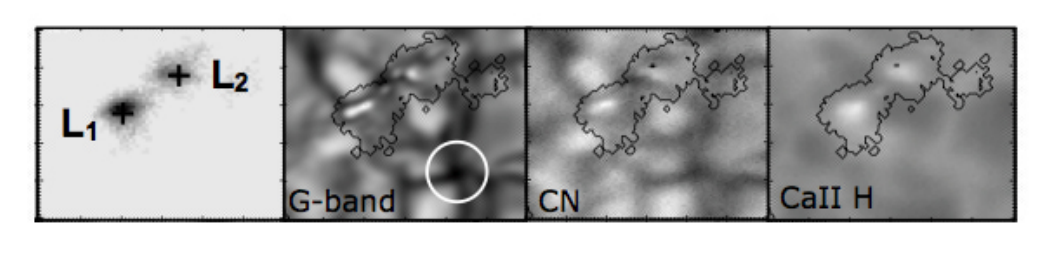} 
\caption{Sample of the observed region on 29 September 2007 at 08:48 UT (4-min average images) combining Hinode and SST observations. \emph{From left to right}:  magnetogram, G-band, CN and Ca \textsc{ii}~H. The FOV is $\sim$6$\arcsec \times~$5$\arcsec$. L1 and L2 are the main two magnetic lobes identified in the region. The 1-Mm diameter white circle is centered in the location where a vortex motion is formed. Black contours outline the magnetic negative polarity region. The FOV corresponds to the region studied in Paper I.}
\label{fig1}
\end{figure}

\begin{figure}
\centering
\includegraphics[angle=0,width=1.\linewidth]{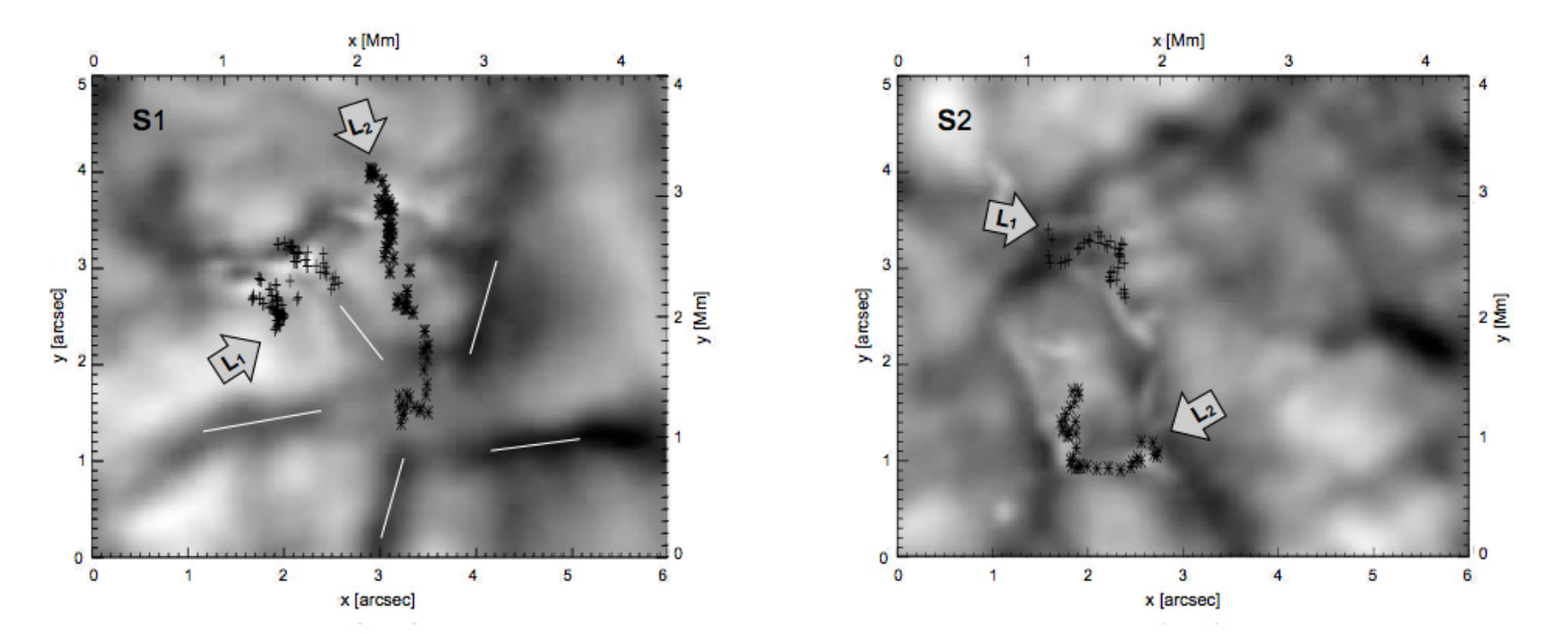}
\caption{Evolution in time of the magnetic centroid for two magnetic lobes L1 and L2. Trajectories followed by the centroids are
independently plotted for both SST/G-band series (S1/S2 for upper/lower panels) respectively. The arrows indicate the starting point for the L1 and L2 trajectories. Backgrounds are the corresponding time average images in every case. Intergranular lanes in the upper panel are outlined in white.}
\label{fig2}
\end{figure}

\begin{figure}
\centering
\includegraphics[angle=0,width=1.\linewidth]{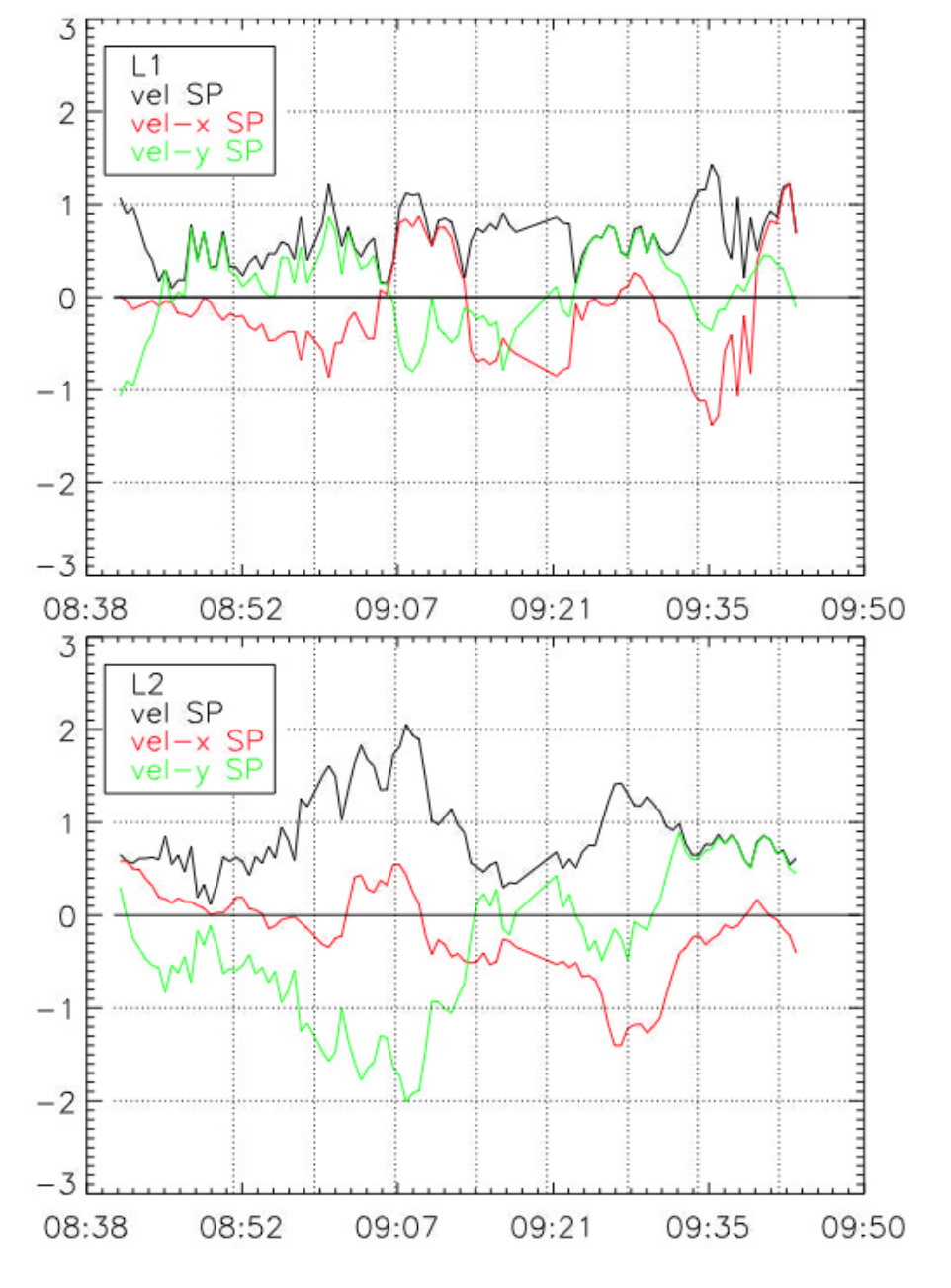}
\caption{Time evolution of horizontal velocities magnitude (\emph{black}), velocity in x (\emph{red}) and velocity in y (\emph{green}) for the magnetic centroid, as computed from the Hinode/SP data for L1 (\emph{upper panel}) and L2  (\emph{lower panel}). Times are in UT and velocities in km s$^{-1}$.}
\label{fig3}
\end{figure}

Two main magnetic lobes (L1 and L2) are identified in Paper I with one of them rotating around the other. Figure~\ref{fig1} shows a sample of images taken at 08:48 UT. Every frame  displays a 4-min average to enhance features. The two magnetic lobes are clearly discernible in the first frame in the figure. In Fig.~\ref{fig2}, we show the trajectories followed by the magnetic centroid for L1 (plus-signs) and L2 (asterisks) calculated from the individual magnetic strength maps obtained from SP data.  Panels in the figure correspond to time intervals for G-band time series \emph{s1} and \emph{s2} (see Table~\ref{table1}), and arrows indicate initial position for every trajectory and magnetic lobe, i.e. 08:47 UT for \emph{s1} and 09:14 UT for \emph{s2}. Trajectories are plotted overlaid to the average image in every case, i.e. 19 and 32 min for series \emph{s1} and  \emph{s2} respectively.  In Fig.~\ref{fig2}(S1) the locations of the centroid for L1 are practically confined in a small area as compared to L2 centroids noticeably drifting towards the lower part of the FOV. Intergranular dark lanes are clearly seen in the average images having widths of the order of 500 km. Straight \emph{white lines} outline the mean location of the intergranular lanes that are found to be converging remarkably into a vertex in the granular pattern (referred to as the \emph{draining point} in Paper I and corresponding to the center of a convective vortex motion).  Moreover, these \emph{white lines} seem to be azimuthally equally-spaced forming an angle of $\sim$58 $\pm$ 4 degrees between two adjacent lines, except for the angle formed by the two lower right ones that is wider. Large horizontal flows coming from the lower right part of the FOV, as will be described in Section~\ref{S:6}, might be causing this angular broadening. The white circle in Fig.~\ref{fig1} is centered in the location where a vortex motion is formed during the corresponding time interval. The region is dominated by an intergranular lane (see arrow L1). Trajectory for L2 changes drastically from the one in (S1) and the converging intergranular lanes are not visible anymore. There are signs of bright features placed along many asterisks in L2. The configuration of the integranular lanes and the evolution of the magnetic centroids in panels (S1) and (S2) in Fig.~\ref{fig2} are proposed to be linked to the presence of a convective vortical motion. Panel (S1) corresponds to the time of appearance of a vortical motion as detected in the plasma flow maps whereas panel (S2) reveals no hints of this type of motion, as will be further detailed in Section~\ref{S:6}. Figure~\ref{fig3} plots the evolution in time of the magnetic centroid horizontal velocity magnitude (\emph{black}), velocity in $x$ (\emph{red}) and $y$ (\emph{green})  directions, as computed from the Hinode/SP data for L1 and L2. The centroid velocities have been computed from the pairs of consecutive centroid positions for the respective lobes.  For L2, we obtained that these velocities increase rapidly from $\sim$0.5 km s$^{-1}$ up to $\sim$2 km s$^{-1}$ during the time interval 08:40--09:07 UT, and then decrease. After that, the velocity increases again, reaching a second maximum of about 1.4 km s$^{-1}$ at around 09:25 UT. Horizontal velocities for L1 show a similar variation in magnitude during the first 30 min. L1 does not undergo significant motion. Towards the end of the sequence, there is an increase in the velocity up to 1.4 km s$^{-1}$ while L1 is dragged by the surrounding flow.

\section{Photospheric plasma motions associated to a swirl event}
\label{S:4}

\begin{figure*}
\centering
\includegraphics[angle=0,width=1.\linewidth]{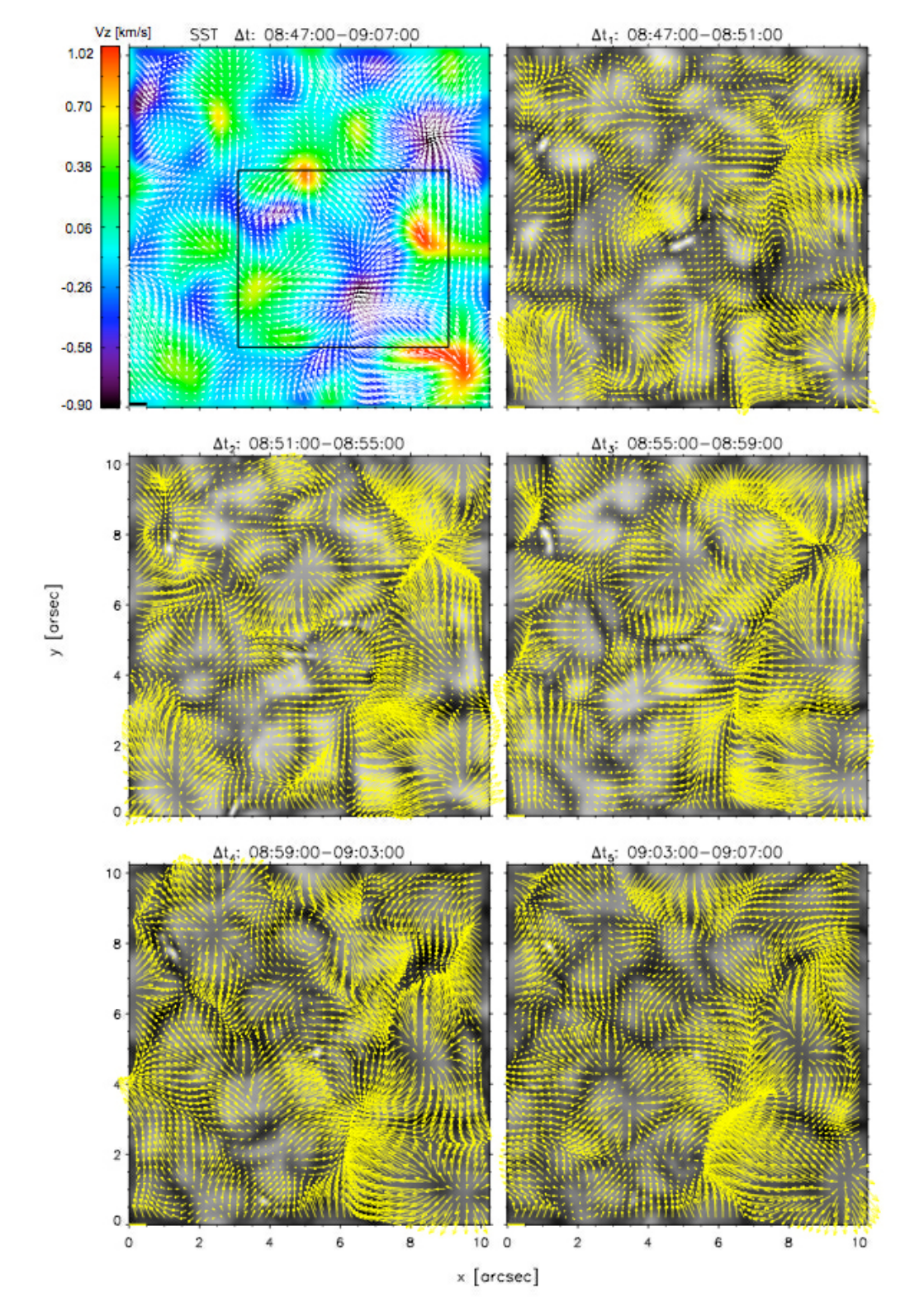} 
\caption{Maps of horizontal velocities for different time intervals covering the  G-band series \emph{s1}. Upper left map is computed for the whole duration of the series against the background image of vertical velocities in false color. The ROI is framed in a black box with the same FOV in Figs.~\ref{fig1} and \ref{fig7}. The other maps are computed over 4-min time intervals as labeled with background images displaying the corresponding average images. The length of the yellow horizontal bar at coordinates [0,0] in every map corresponds to  1.5 km s$^{-1}$.}
\label{fig4}
\end{figure*}

Solar plasma at the photospheric level displays a rapid evolution of the order of a few seconds. The high-cadence of the acquired time series enables to follow the evolution of the granulation pattern. In this section we aim to characterize the photospheric flows in the ROI displaying the convective swirl event. The SST/G-band and Hinode/CN time series have been independently employed to compute the horizontal proper motions of structures by means of a local correlation tracking (LCT) technique \citep{november1988} implemented by \cite{molowny1994}.  Maps of horizontal velocities are calculated for time series \emph{s1} and \emph{s2} using a Gaussian tracking window of FWHM $1.0 \arcsec$ (roughly half of the typical granular size).  We applied the LCT analysis over a FOV of $\sim$10$\arcsec$$\times$10$\arcsec$ that includes the ROI (of 6$\arcsec$$\times~$5$\arcsec$ shown in Figs.~\ref{fig1}, \ref{fig2} and \ref{fig7}). The flow maps computed from SST time series \emph{s1} are displayed in Fig.~\ref{fig4}, where the black box in the colored map represents the ROI FOV. The map in the upper-left panel is calculated by averaging over the total duration of the series ($\Delta t$=19 min). The underlaying background represents the image of vertical velocities computed from the divergences of the horizontal velocity field following \cite{november1988}. Flows coming from granular explosive events are dominant in the FOV and are normally associated with mesogranulation \citep{roudier2004,bonet2005}. These events are clearly seen in the image of vertical velocities as strong positive magnitudes (upflows). We only identified  two cases of strong sinks, e.g. negative vertical velocities with magnitude  $\sim$0.9 km s$^{-1}$ (downflows), displaying converging horizontal flows (i.e. the velocity arrows point to a common origin) situated along intergranular lanes at coordinates [6.5,3.5] and [8.5,7.5], respectively. The first case corresponds to the main vortex we are interested in, that is located within the ROI studied in this work, and the second case, referred to as V2, will only be briefly commented in this section since it does not seem to affect the dynamics of the small magnetic elements in the region.

The statistics of the horizontal velocities are computed for all the time intervals ($\Delta t$ and $\Delta t_1$ to $\Delta t_5$). Figure~\ref{fig6} plots the histograms where the black represents the one for $\Delta t$ and the color ones correspond to the 4-min intervals averages. The velocity statistics are computed in a region including the vortex by establishing a threshold in the divergence field to mask the vortex area coverage thus computing the vectors in the nearest vicinity from the centre of convergence. The small upper panel in the figure plots the mean velocity magnitude for the 4-min intervals as colored-diamonds (same colors as the ones used for histograms) in the same vortex areas. The horizontal \emph{dotted-line} ($\Delta t_{vortex}$) represents the value for the mean velocity magnitude as computed from the whole time interval $\Delta t$ in the black box framing the vortex (see top left panel in Fig.~\ref{fig4}), and the horizontal \emph{black line} ($\Delta t_{FOV}$) accounts for the same value in the whole FOV ($\sim10\arcsec\times10\arcsec$).  There is a clear increasing trend for the mean velocity magnitudes  when moving in time from $\Delta t_1$ to $\Delta t_5$, starting from a value of $\sim0.7$ km s$^{-1}$ ($\Delta t_1$) very close to the mean velocity magnitude averaged over the whole FOV ($\Delta t_{vortex}$; when the vortex has not yet been formed). A period of stability (of about 10 min) in the velocity magnitude is registered for  $\Delta t_3$ and $\Delta t_4$ corresponding to steady-vortex flows in the maps in Fig.~\ref{fig4}. After that, for interval $\Delta t_5$, we evidence large flows as coming from a strong explosive event taking place at the lower right corner of the FOV, that seem to be destroying the steady configuration of the vortex by sweeping away weaker velocity vectors. 

To be able to establish a value for the lifetime of the vortex, the LCT is extended to later temporal intervals by using the second time series (series \emph{s2} in SST, Table~\ref{table1}). Unfortunately we have a 6-min gap where there is not available data. The map of horizontal proper motions averaged over the whole duration of  \emph{s2} (32 min) is displayed in the top left panel in Fig.~\ref{fig5} with the background representing again the computed vertical velocities (note that the scale is not comparable to the one in  Fig.~\ref{fig4} since the averages are computed over time series of different durations, i.e. 19 and 32 min respectively, as commented above). The black box extracts the same ROI as in the upper-left panel in Fig.~\ref{fig4}. The once strong sink (and vortex) in the analysis of series \emph{s1} has turned into a smoother structure with downward velocity magnitudes of only a few km per sec in magnitude ($\sim$2 km s$^{-1}$) in our velocity scale. In an attempt to identify whether the vortex has completely disappeared for shorter time averages, we pursued the same procedure as before by averaging over 4-min intervals. The flow maps in Fig.~\ref{fig5} display no signatures of this steady photospheric vortex-type flow. Only for the first 4-min average map there is a small sign of what could be the remaining of a diluted vortex shifted to the left by $\sim$1 arcsec as pushed away by strong velocities starting to develop in lower right part of the FOV in the last panel in Fig.~\ref{fig4}. According to this analysis, we estimate the lifetime of the vortex to be in the range between 15 to 19 min. These values are larger than the $\sim$5 min lifetime of the vortex-type motions described by e.g. \cite{bonet2008} but, different from the calculation of these authors, we have not estimated it depending on the evolution of BPs in the region but from the proper motions of convective plasma. 

Though our studied vortex seems to have disappeared in the flow maps for \emph{s2}, the other important sink in the FOV hosting vortex V2 remains present and in a developed stage. The evolution of V2 can be actually followed throughout the flow maps in Figs.~\ref{fig4} and \ref{fig5} at coordinates (8, 7.5).  Vortical motion appears to be more stable and lasts longer but with a similar spatial coverage.  Surprisingly, this vortex does not seem to be affecting the dynamics of BPs in their vicinity in the G-band time sequence. Concerning Hinode data, the map of horizontal velocities is computed over the CN sequence (not shown in the paper) and the results are comparable to the ones described above for SST/G-band data, though the noise increases up due to the lower spatial resolution.

\section{Activity of small magnetic elements}
\label{S:5}

The studied quiet Sun region is populated by BPs associated to the magnetic concentrations, as commented in previous sections. The activity of BPs reveals intrinsic properties of the intense magnetic fields that emerge from the solar interior at such small spatial scales \citep[e.g.][]{ishikawa2007}  . Using the high-resolution G-band time series we can identify the rapid evolution of the region, i.e. exploding granules, reorganization of the granular pattern and a population of BPs. Figure~\ref{fig7} displays the sequence of images  of the ROI (including the two SST/G-band time series \emph{s1} and \emph{s2} in false color, separated by the grey-scaled Hinode/CN image at 09:12 UT. The white arrow is a reference pointing to the location of the  swirl convective motion that develops in the region. Single BPs and chains of BPs are highly dynamic and we can visually evidenced clear stages of fragmentation (i.e. 08:48--09:00 UT) and coalescence (09:05--09:20 UT) accompanied by spatial displacements and intensity changes.

The intensity variations in the series of CN, Ca \textsc{ii}~H, G-band and magnetograms are computed in a small box  (7 $\times$ 7 square pixels) at selected points (colored dots in first panel in Fig.~\ref{fig7}. Figure~\ref{fig8} plots the variation in time of intensity for two of the selected locations (\emph{black} and \emph{blue} dots) in the vortex's nearest vicinity. Although special attention is drawn to the events of convective collapse occurring after 09:20 UT that are described in the next section, we include their intensity profiles for the whole time range. From the analysis of these profiles together with visual exploration of Fig.~\ref{fig7} we can infer that the region undergoes significant changes due to the vortex motions in the early phase. Later on, important peaks in G-band intensity are seen at around 09:20--09:25 (black curve) and 09:32--09:40 (blue curve). Peaks in the intensity profiles are associated to the formation of BPs product of coalescence processes and they accompany changes in the underlying magnetic field as a consequence of the vortex motion described in the previous section. The magnetic activity and average intensity in some locations are strongly linked to the rotation of the magnetic structures that seems to be dragged by the vortex developed during the time coverage of \emph{s1}.  Some of the flows  in Fig.~\ref{fig4} are precisely observed within the ROI along trajectories connecting BPs and ending up in the region where the center of the vortex is located, and hence bringing plasma and embedded magnetic elements together to the vortex.

\subsection{Magnetic field intensification}
\label{collapsesec}

Using Hinode SP data, we analyze two particular cases for plausible convective collapses taking place close to the vortex region. Magnetic field magnification/intensification is observed soon after the vortex has disappeared.  Figure~\ref{fig9} shows: {\bf 1)} Stokes $V$ profiles of the magnetically sensitive 6301 and 6302~\AA~ Fe~$\textsc{i}$ lines, {\bf 2)} maps displaying LOS velocities (V$_{LOS}$) obtained by Dopplershift calculation on Stokes $I$ and {\bf 3)} co-temporal magnetograms in absolute value  ($\arrowvert V \arrowvert / \langle I_{c}\rangle$) obtained from the peak Stokes V. The two cases are referred to as Case~1 and Case~2. In Case~1 (see row labeled A in Fig.~\ref{fig9}) maps are displayed every 32 s while Stokes $V$ profiles are plotted every 66~s (i.e., once every two maps)  measured at the centre of the region encircled in white in the maps (at 09:22 UT). A green crosshair pinpoints the area where these profiles are obtained. Asymmetries between the two lobes are evident. In the fourth panel in A, a three-lobed profile appears in the 6301~\AA~but not in the 6302~\AA~line, suggesting the presence of stronger gradients of magnetic field strength or velocity higher up than in the lower photosphere (6301~\AA~line). This phenomenon occurs during the period 09:22--09:25 UT, when the maximum LOS velocity is about -3.5~km s$^{-1}$.  The sequence in row B  (Case~1) in the same figure shows Stokes $V$ profiles for an adjacent pixel, pinpointed in yellow by a crosshair. Again, the appearance of three-lobed Stokes $V$ profiles indicate gradients of velocity. In the fourth panel, a sudden appearance of a more symmetric profile suggests 'stabilization' of the magnetic field or velocity gradients of the BP. In the G-band filtergrams, the presence of new BPs associated to this event are observed in this region (see encircled region in frame at 09:20 UT in Fig.~\ref{fig7}). For Case~2, the symmetric profiles, calculated for the encircled region in the corresponding maps at 09:34 and the pixel marked by a green crosshair, do not suggest strong changes on gradients and the magnetic region is rather stable in time. However, during an interval of 66~s we observe a variation in velocity from -1.0 to -3.5~km s$^{-1}$. Similar to Case~1, BPs appear in this location in the G-band images (encircled in white at 09:34) simultaneously to the convective collapse event  and lasting for about 8 min. 

Velocities in the magnetized atmosphere can be estimated from a zero-crossing Stokes $V$. We can perform a rough guess of the velocity values from the Stokes $V$ profiles and their nominal wavelengths, marked with magenta for the 6301~\AA~and blue for the 6302~\AA~line. We estimate the wavelength shifts are about 0.1-0.15 \AA, which lead to downflow velocities in the range of -4.7 to -7.1 km s$^{-1}$. Therefore, another key element of the convective collapse, the supersonic downflow, appears into play.
 Both cases extend over an area of about 4--6 pixels (0\farcs30 $\times$ 0\farcs30) and have lifetimes of the order of 5--8 min.  Asymmetric profiles and downflows have been found and explained in different works, e.g. in \citet{Orozco_2008_granules, fischer2009}. Also \citet{Bello_2009_simulgbp} found asymmetric Stokes $V$ profiles where a BP showed flux intensification and downflows. 

\section{Discussion}
\label{S:6}

\begin{figure*}
\centering
\includegraphics[angle=0,width=1.\linewidth]{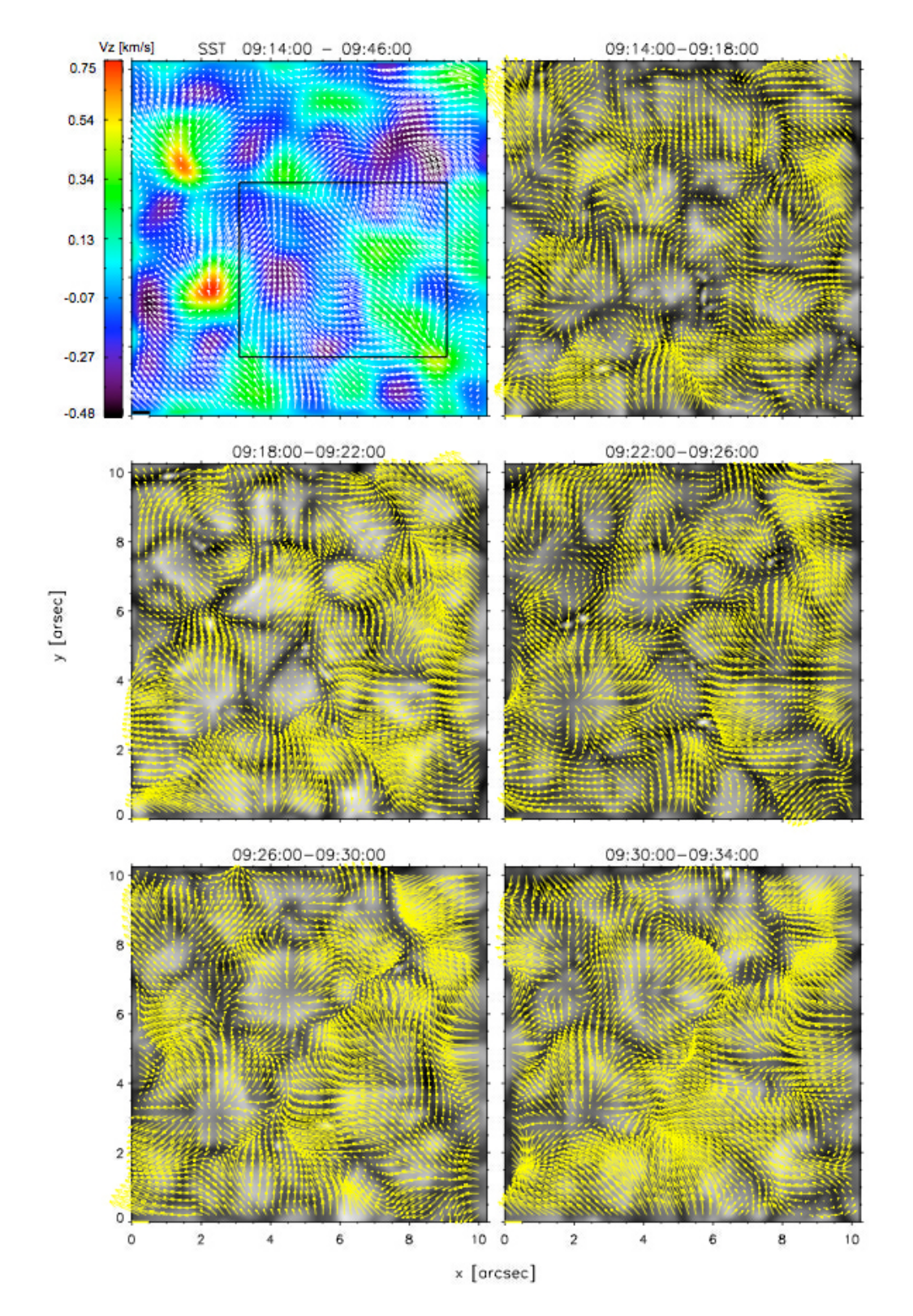} 
\caption{The same as in Fig.~\ref{fig4} computed over  G-band series \emph{s2}. Note that the scale for vertical velocities differs from the one in Fig.~\ref{fig4} due to the 
different duration of the time series affecting the velocity averages. See the text for details.}
\label{fig5}
\end{figure*}

\begin{figure}
\centering
\includegraphics[angle=0,width=1.\linewidth]{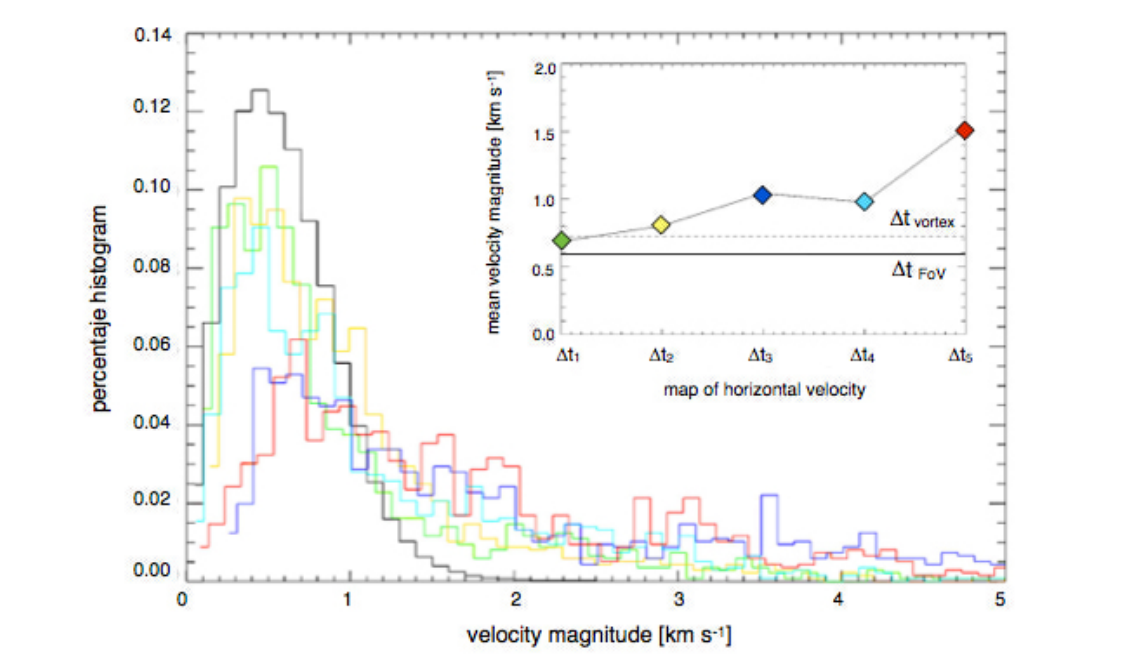} 
\caption{Histogram of horizontal velocities computed for different time intervals for G-band time series \emph{s1}, $\Delta t$=19 min (in \emph{black}) and 4-min intervals averages $\Delta t_1$ to $\Delta t_5$ (see the small upper panel for the corresponding colors). Statistics of velocities are computed in a region including the vortex. The upper panel plots the mean velocity magnitudes in the histogram for the different interval, where $\Delta t_{vortex}$ (\emph{dotted line}) and $\Delta t_{FOV}$ (\emph{black line}) represent the mean velocity computed in the box framing the vortex and in the whole FOV in the top left panel in Fig.~\ref{fig4}, respectively (see text for details).}
\label{fig6}
\end{figure}

\begin{figure*}
\centering
\includegraphics[angle=0,width=1.\linewidth]{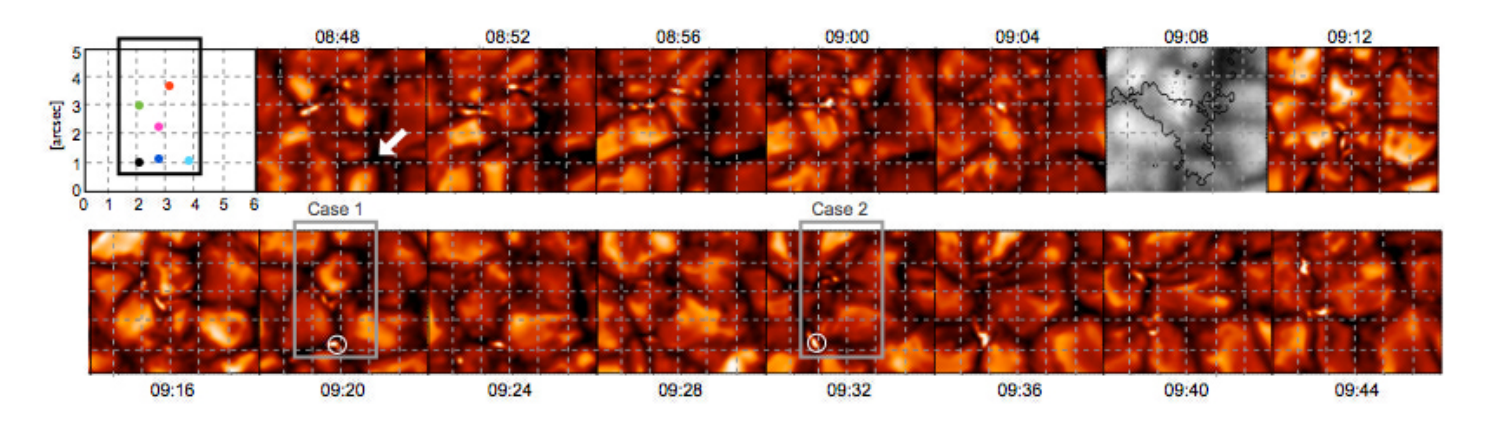}   
\caption{Sequence of contrasted SST/G-band images for both time series \emph{s1} and \emph{s2} showing the evolution of the studied quiet Sun region on 29 September 2007. Frame at 09:08 UT (in grey-scale) corresponds to a Hinode/CN image that covers the gap between \emph{s1} and \emph{s2} during the SST observation, with black contours outlining the negative magnetic polarity area. Every frame displays the average image over 4-min intervals and the time stamps correspond to the initial time.  The top left frame shows the location of  5 regions of interest (\emph{colored dots}). The white arrow in frame at 08:48 pointing to the center of a convective vortex motion detected in the FOV is included for reference.  Boxes in black (top sequence) and grey (bottom sequence) extract a common FOV. Case~1 (at 09:20) and Case~2 (at 09:32)  stand for the analysis of convective collapse events shown in Fig.~\ref{fig9}. The locations displaying an intensification of BPs  are encircled in white.} 
\label{fig7}
\end{figure*}

\begin{figure*}
\includegraphics[angle=0,width=1.\linewidth]{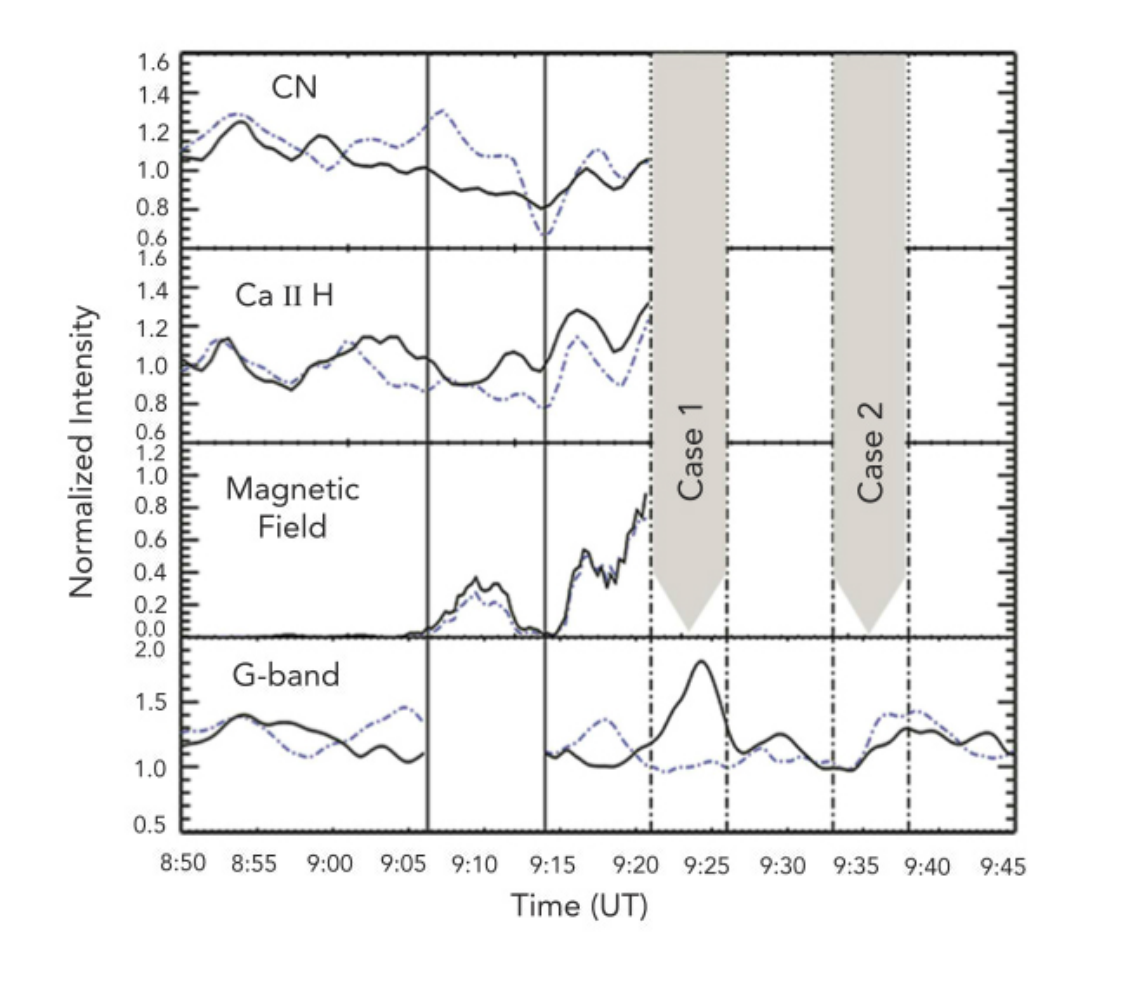} 
\caption{Normalized intensity profiles for selected locations in different series of images. First three rows (from top to bottom) from the analysis of Hinode/BFI (CN and Ca \textsc{ii}~H)  and NFI (magnetic field) data, and the last row from SST data. The vertical solid lines separate the two time series \emph{s1} and \emph{s2} in SST/G-band data with the 6-min gap in between. Blue and black curves correspond to the locations of corresponding \emph{colored dots} in the top left panel in Fig.~\ref{fig7}, respectively.  The two vertical shadowed areas highlight the time periods of the plausible convective collapse events (Cases 1 and 2) analyzed in Fig.~\ref{fig9} that correspond to the location of the \emph{dark-blue} and \emph{black dots} in the first panel in Fig.~\ref{fig7} and encircled in white in the same figure.}
\label{fig8}
\end{figure*}

In the ROI (10 $\times$ 10 square arcsec) analyzed in the present work  a couple of examples of convective vortical motions (affecting a circular area of radius $\sim$3.0\arcsec from the vortex center) are found with one of them lasting for more than an hour. The stability and lifetime of vortical motions are strongly linked to the evolution of the granular pattern and they seem to gather larger velocity magnitudes in their more developed stage. Paper I evidenced small-scale magnetic concentrations of negative polarity affected by one of these convective swirl motion of plasma, and identified two main lobes of which one ot them does not seem to be affected and remains rather still while the other lobe is dragged an approaches the center of the vortex. In our analysis of the same region we determined the motion of the magnetic centroid for each of these lobes and show that one of the lobes is affected as long as the vortex is present. The configuration of the integranular lanes with an azimuthally quasi-symmetric distribution as seen in the average image and the evolution of the magnetic centroids are proposed to be linked to the presence of a convective vortical motion. The vortical motion has different developing stages as evidenced by the computed horizontal flow maps showing a more stable period lasting for about 10 min.  The onset of the vortex is characterized by rather low mean velocity magnitudes (600  m s$^{-1}$) and the value increases up to 1 km s$^{-1}$ while reaching the stability period mentioned above. The fate of the vortex seems to be determined by strong flows coming from the lower left part or FOV that sweep away the coherence of the vortical motions and increase the mean velocity value to about 1.5  km s$^{-1}$. We estimate the lifetime of the vortex to be $\sim$15 min. The configuration of the region in terms of horizontal flows, after the vortex has disappeared, resembles the one for a mesogranular flow pattern (see for instance the lower panels in Fig.~\ref{fig5}). In the work by \citet{bonet2008} BPs are found to describe spiral trajectories towards the centre of convectively driven vortex flows, though these authors do not claim to observe fragmentation or coalescence processes associated to the activity of the visually tracked BPs. In our analysis we found intense activity of BP«s including fragmentation and intensification due to coalescence at the time when the magnetic concentrations rotate and seem to be dragged by the formation of the convective vortex motion (08:48 to 09:08 UT in the sequence in Fig.~\ref{fig7}).  The comparison with the activity of BPs in regions isolated from the effects of convective vortex motions is out of the scope of the present work, but we acknowledge that should be properly addressed to determine the complete influence of these types of motions in shaping the granular patterns at fine spatial scales..

\begin{figure*}
\includegraphics[angle=0,width=1.\linewidth]{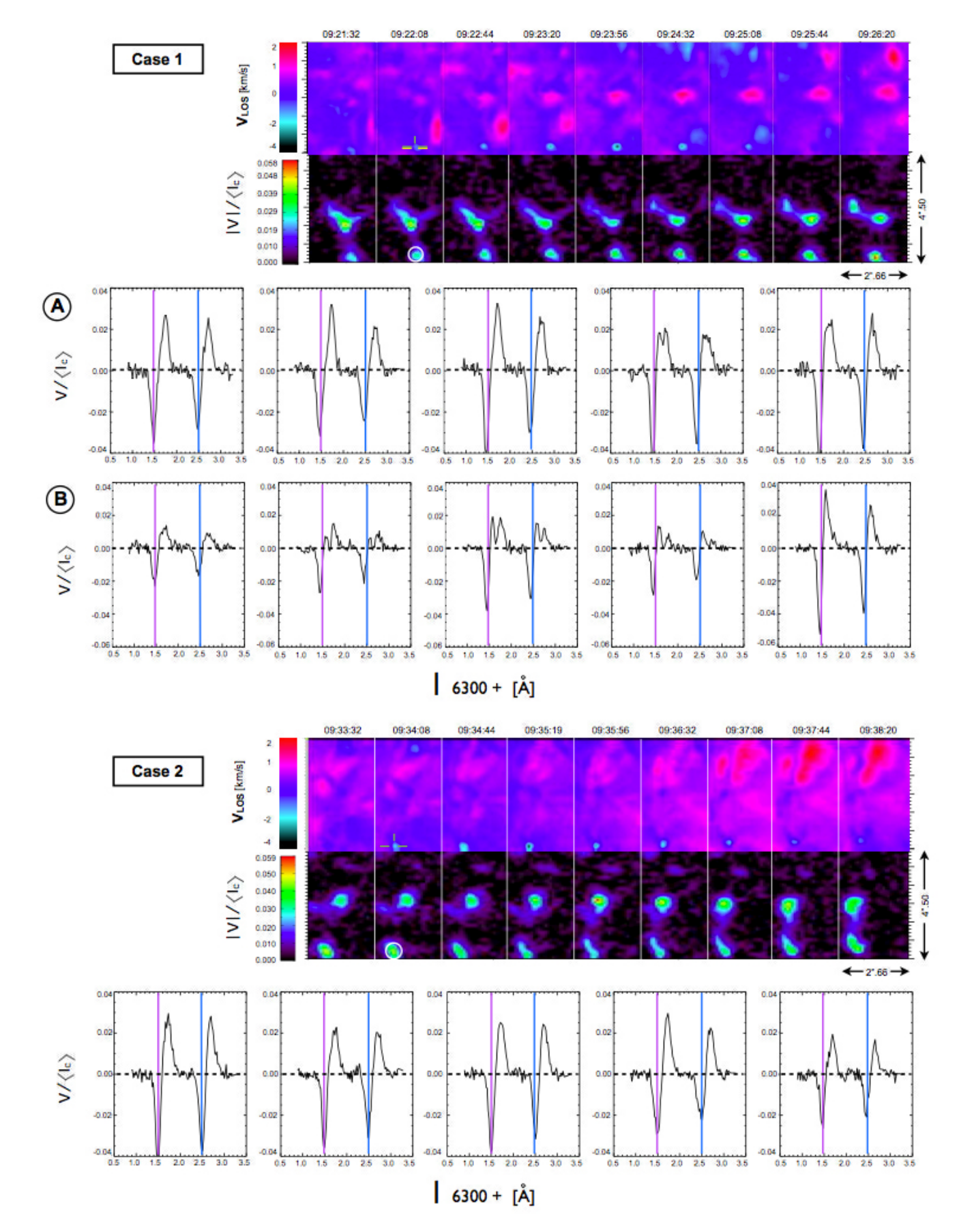} 
\caption[Velocity and magnetic field maps for two convective collapses]{Detected cases of plausible convective collapse events in Hinode data on 2007, September 29. For every case the temporal sequence of LOS Doppler velocities ({\it upper row}) and magnetic signal ({\it lower row}) are simultaneously shown in false-color maps. Encircled in white are regions displaying redshifts accompanied by an intensification in the magnetic field. Stokes $V$ profiles are correspondingly plotted for these regions (from the subpanel holding the white circumference to the following four ones), along with magenta and blue lines indicating the nominal wavelengths.}
\label{fig9}
\end{figure*}

In the cases presented in Figure~\ref{fig9}, small magnetic patches experience a sudden strong downflow and some intensification of the magnetic field strength. Since magnetic elements are actually pres1ent in the locations before the event (i.e. the strong downflow appearance), these cases are likely convective collapses of a part of the magnetic area or a flux intensification, and could be explained as unstable flux tubes with high field strength \citep{spruit1979}. According to \citet{bellot2001}, the process of flux concentration is connected with strong redshifts, while the destruction of a magnetic structure is associated with large blueshifts. In \citet{Narayan_2011}, the increasing velocities in downflows correspond to magnetic field intensification. These cases are interesting, since they can provide an observational counterpart to the simulations, for instance, of \citet{danilovic2010}.  Apart from the studied vortex, there is another vortex (V2) in our region of interest, that is very stable and last for at least 40 min. An interesting fact is that the BPs detected at the beginning of the time series are actually closer in space to V2 than to the vortex that is dragging the magnetic concentrations. Magnetic elements seems not to be affected by V2. In order to understand this scenario we would require to determine how the flux tubes associated to the BPs are rooted and ascertain possible connections and interactions that might be influenced by the convective flows below the photospheric level.  As far as our observational evidence is concerned, the small-scale flux tubes can be deformed underneath the optical surface with a region closer to the vortex that seems to affect them yet they could be located  closer to other vortex (V2 in our studied example) at the photosphere.
 
 The evolution of photospheric flows at different spatial scales might activate solar activity in upper layers by triggering reconnection and propagation of waves, among others, e.g. supergranular flows, as reported by \cite{innes2008}. Small-scale vortical motions on the solar photosphere are therefore thought to play an important role in the dynamics of the quiet Sun \citep{ballegooijen1998,kitiashvili2011}, not only by merging magnetic concentrations thus forming more stable structures (i.e. pores) but also by promoting displacement of  footpoints  associated to small magnetic loops connecting the surface and the upper chromospheric layer hence allowing reconnection to take place when opposite polarity magnetic fields are present. 

According to simulations, photospheric vortical motions contribute merging small magnetic concentrations, i.e. the magnetic field is advected by the flow, as commented above, and this a plausible mechanism responsible for the formation and evolution of stronger magnetic concentrations and ultimately pores, accompanied by strong downflows down to a distance of a few Mm beneath the visible solar surface. The analysis of plasma motions around solar pores shows that their nearest vicinity is dominated by inflows  \citep{vargasthesis,vargas2010} and moreover, signs of downflows have also been detected in regions around pores \citep{hirzberger2003}. More observational evidence should be studied to figure out the real connection and the acting part of convective vortex flows in this context.  In our studied region there are only negative polarity elements but there are also some evidence \citep{kubo2010} claiming the cancellation of opposite-polarity magnetic elements as approaching a junction of intergranular lanes in regions displaying converging horizontal flows.  

Converging and downward motions have therefore proved to be important bringing together magnetic elements either with the same or with opposite polarities and changing the topology of small low-lying loops in the solar surface. Convective vortex-type motions should be considered contributing into this scenario to determine the way they affect the evolution of magnetic elements from the photosphere to upper layers, therefore the importance of the presented observational findings adding new evidence on the dynamics of small-scale structures in the solar photosphere.

\begin{acks}
JP acknowledges funding from the spanish grant BES-2007-16584 and PPII10-0183-7802 from the Junta de Comunidades de Castilla-La Mancha of Spain. JP, VD and IC acknowledges funding from  the projects ESP2006-13030-C06-04 and AYA2009-14105-C06, including European FEDER funds. The Swedish 1-m Solar Telescope is operated on the island of La Palma by the Institute of Solar Physics of the Royal Swedish Academy of Sciences in the Spanish Observatorio del Roque de los Muchachos of the Instituto de Astrof\'isica de Canarias. We thank the scientist of the Hinode team for the operation of the instruments. Hinode is a Japanese mission developed and launched by ISAS/JAXA, with NAOJ as domestic partner and NASA and STFC (UK) as international partners. It is operated by these agencies in co-operation with ESA and NSC (Norway).
\end{acks}


\newpage


\end{article} 

\end{document}